\documentclass[a4paper]{article}

\usepackage{amsmath}
\usepackage{amssymb}
\usepackage{amsthm}
\usepackage{amsfonts}

\newtheorem{theorem}{Theorem}

\theoremstyle{definition}

\newcommand{\CC}{\mathbb{C}} 
\newcommand{\ZZ}{\mathbb{Z}} 

\DeclareMathOperator{\im}{Im} 
\DeclareMathOperator{\diag}{diag} 
\DeclareMathOperator{\Sp}{Sp} 
\DeclareMathOperator{\Sing}{Sing} 
\DeclareMathOperator{\mult}{mult} 

\newcommand{\smsum}{\mathop{\textstyle{\sum}}\limits} 

\newcommand{\A}{\mathcal{A}} 
\newcommand{\B}{\mathcal{B}} 
\newcommand{\II}{\mathbb{I}} 
\newcommand{\M}{{\mathcal M}} 
\newcommand{\Sieg}{\mathfrak{H}} 

\newcommand{\tp}{\,{}^t\!} 

\newcommand{\OO}{\mathcal{O}}

\newcommand{\spin}[2]{\Bigl[\begin{matrix}#1\\[-4pt]#2\end{matrix}\Bigr]}

\title{Superstring measure and non-renormalization of the three-point amplitude}
\author{Marco Matone and Roberto Volpato}\date{}

\begin{document}
\maketitle

\begin{center}Dipartimento di Fisica ``G. Galilei'' and Istituto
Nazionale di Fisica Nucleare \\
Universit\`a di Padova, Via Marzolo, 8 -- 35131 Padova,
Italy\end{center}

\bigskip

\begin{abstract} We show that a recently conjectured
expression for the superstring three-point amplitude, in the
framework of the Cacciatori, Dalla Piazza, van Geemen -- Grushevsky
ansatz for the chiral measure, fails to vanish at three-loop, in
contrast with expectations from non-renormalization theorems. Based
on analogous two-loop computations, we discuss the possibility of a
non-trivial correction to the amplitude and propose a natural
candidate for such a contribution. Thanks to a new remarkable
identity, it is reasonable to expect that the corrected three-point
amplitude vanishes at three-loop, recovering the agreement with
non-renormalization theorems.
\end{abstract}

\bigskip

In the last years there has been a considerable progress in the
conceptual understanding and in derivation of explicit formulas for
multiloop superstring amplitudes. In a series of papers
\cite{D'Hoker:2001zp,D'Hoker:2004xh,D'Hoker:2004ce,D'Hoker:2005jc,D'Hoker:2005ia,D'Hoker:2007ui},
D'Hoker and Phong have derived from first principles an explicitly
gauge independent expression for the $2$-loop chiral superstring
measure on the moduli space of Riemann surfaces, given by
\begin{equation}\label{twoloopmeas}d\mu[\delta]=\theta[\delta](0)^4\Xi_6[\delta]d\mu_{Bos}\ ,
\end{equation}
where $\delta\in\ZZ_2^2$ is an even spin structure, $\Xi_6[\delta]$
is a modular form of weight $6$ for the subgroup
$\Gamma(2)\subset\Sp(4,\ZZ)$ leaving theta characteristics invariant
and $d\mu_{Bos}$ is the (genus $2$) bosonic string measure.
Analogous procedures also led them to prove the
non-renor\-ma\-liza\-tion of the cosmological constant and of the
$n$-point functions, $n\le 3$, up to $2$-loops, as expected by
space-time supersymmetry arguments \cite{Martinec:1986wa}.
Furthermore, the $4$-point amplitude has been computed and checked
against the constraints coming from S-duality \cite{D'Hoker:2005ht}.

Direct computations of higher loop corrections to superstring
amplitudes appear, at the moment, out of reach. However, the strong
constraints coming from modular invariance and from factorization
under degeneration limits, together with the explicit $2$-loop
expressions, can lead to reliable conjectures on such corrections.
This is the point of view adopted, for example, in
\cite{Matone:2005vm}, where a general formula for the higher loop
contributions to the $4$-point function has been proposed.

In \cite{D'Hoker:2004xh,D'Hoker:2004ce}, D'Hoker and Phong
conjectured that the $3$-loop chiral superstring measure may be
expressed in the same form as \eqref{twoloopmeas} for a suitable
modular form $\Xi_6[\delta]$ of weight $6$. Such a form is required
to fulfill a series of constraints related to holomorphicity,
modular invariance and factorization; however, no form has been
found satisfying all such conditions. In \cite{Cacciatori:2008ay}
Cacciatori, Dalla Piazza and van Geemen (CDvG) showed that a series
of equivalent constraints can be solved by assuming the more general
expression
\begin{equation}\label{threeloopmeas}
d\mu[\delta]=\Xi_8[\delta]d\mu_{Bos}\ ,
\end{equation}
for the $3$-loop chiral measure, where $\Xi_8[\delta]$ is a modular
form of weight $8$ (not necessarily divisible by
$\theta[\delta](0)^4$). In \cite{DallaPiazza:2008}, it has been
proved that such constraints admit a unique solution for genus $3$,
thus ruling out the more restrictive assumption by D'Hoker and
Phong. Hereafter, we will drop the subscript in $\Xi_8[\delta]$.

The CDvG ansatz for the $3$-loop measure has been generalized by
Grushevsky \cite{Grushevsky:2008zm} to a formula defined for every
loop. In particular, it has been shown that the D'Hoker and Phong
formula for genus $2$ and the CDvG ansatz can be re-expressed in
terms of modular forms associated to isotropic spaces of theta
characteristics, and this leads to a straightforward generalization
of $\Xi$, solving the constraints for all genera. A possible problem
in Grushevsky construction concerns the existence of holomorphic
roots of modular forms appearing in the definition of $\Xi$ for
genus $g\ge 5$. Proving that such roots are well-defined seems
highly non-trivial and, up to now, it has been shown only for genus
$5$ \cite{SalvatiManni:2008qa}.

\medskip

Further consistency checks for chiral measure ans\"atze may be
provided by the non-renormalization of the cosmological constant and
of the $1$,$2$,$3$-point functions. In this respect, it is useful to
consider more closely the general form of the $g$-loop contribution
to the $n$-point function in superstring theories. Using the
notation of \cite{D'Hoker:2007ui}, the general $n$-point amplitude
can be expressed as
$$A(k_1,\ldots,k_n,\epsilon_1,\ldots,\epsilon_n)=\int\prod
dp^\mu_j\int_{\M_g}\int_{C^n}\Bigl|\sum_{\delta\text{
even}}\B[\delta](k_i,\epsilon_i,z_i,p^\mu_j)\Bigr|^2\ ,
$$
where $k_i,\epsilon_i$, $i=1,\ldots,n$, are the space-time momenta
and polarizations of the external states, $p_j^\mu$, $j=1,\ldots,g$,
are the internal loop momenta, $\M_g$ is the moduli space of Riemann
surfaces of genus $g$, $C$ is the Riemann surface at the
corresponding point of $\M_g$ and $z_i$, $i=1,\ldots,n$ are the
insertion position of the vertex operators on $C$. The sum is over
all the even spin structures $\delta$ and
 $\B[\delta](k_i,\epsilon_i,z_i,p^\mu_j)$ are the so-called chiral amplitudes.
In principle, upon applying a consistent gauge fixing procedure, the
chiral amplitudes $\B[\delta]$ can be obtained by computing suitable
CFT correlators of products of vertex operators, supercurrents and
stress-energy tensor operators \cite{D'Hoker:2005jc}. It is often
useful to split the chiral amplitudes into a sum
$$\B[\delta]:=\B^c[\delta]+\B^d[\delta]\ .$$ Here, the so-called
disconnected part $\B^d[\delta]$ includes all the terms where the
Wick contractions between the vertex operators are disconnected from
the contractions of the supercurrents and of the stress-energy
tensor; the connected part $\B^c[\delta]$ includes all the other
terms \cite{D'Hoker:2005jc}.

It is worth noticing that $\B^c[\delta]$ is a sum of a huge number
of terms and that, in general, each of these terms strongly depends
on the details of the gauge fixing; only the whole sum is gauge
independent. For the $2$-loop $n$-point functions with $n\le 3$
(but, for example, not for $n=4$), all the different contributions
to $\sum_\delta\B^c[\delta]$ simply cancel each other, just giving
$\sum_\delta\B^c[\delta]=0$. It follows that the non-renormalization
theorems at $2$-loops are implemented by $\sum_\delta\B^c[\delta]$
and $\sum_\delta\B^d[\delta]$ vanishing separately. For arbitrary
genus $g>2$, a general consistent gauge fixing procedure is not
known (even though several steps of the genus $2$ derivation
generalize to genus $3$). As a consequence, the precise mechanism
for the cancelation of the gauge fixing ambiguities among such terms
is not understood in full detail. It is reasonable, therefore, to
assume that, in analogy with the genus $2$ case, all the ambiguous
terms exactly cancel giving
\begin{equation}\label{connvanish}\sum_{\delta\text{ even}}\B^c[\delta]=0\ ,\end{equation}
for $n\le 3$ and for arbitrary genus.

Under the assumption \eqref{connvanish}, the non-renormalization
theorems are equivalent to the vanishing of the disconnected part
$\B^d[\delta]$ of the chiral amplitude, which can be easily
expressed in terms of the chiral superstring measure. In fact, given
the ansatz \eqref{threeloopmeas}, it can be proved that the
vanishing of the $n$-point function is equivalent to
\begin{equation}\label{Npoint}A_n(z_1,\ldots,z_n):=
\sum_{\delta\text{ even}} \Xi[\delta]\A[\delta](z_1,\ldots,z_n)=0\ ,
\end{equation}
where $\A[\delta](z_1,\ldots,z_n)$ can be derived from the Wick
contractions of the $n$ vertex operators, inserted at the points
$z_1,\ldots,z_n$ of the Riemann surface. More precisely, the
$0$-point function is obtained by setting $\A[\delta]=1$, so that
the non-renormalization theorem for the cosmological constant is
just equivalent to
\begin{equation}\label{zeropvanish}A_0\equiv\sum_{\delta\text{ even}}\Xi[\delta]=0\ .
\end{equation} Notice that \eqref{connvanish}
trivially holds for $n=0$; the vanishing of the cosmological
constant is, therefore, the most reliable check of the correctness
of the chiral superstring measure ans\"atze. The $1$-point function
vanishes automatically under the assumption \eqref{connvanish},
whereas the non-renormalization of the $2$- and $3$-point amplitudes
corresponds to
\begin{align}
A_2(a,b)&\equiv \sum_{\delta\text{ even}}\Xi[\delta]S_\delta(a,b)^2=0\ ,\label{twopvanish}\\
A_3(a,b,c)&\equiv \sum_{\delta\text{
even}}\Xi[\delta]S_\delta(a,b)S_\delta(b,c)S_\delta(c,a)=0\label{threepvanish}\
,
\end{align}
respectively, where $a,b,c$ are arbitrary points of the genus $g$
Riemann surface $C$ and $S_\delta(a,b)$ is the Szeg\"o kernel. The
identity \eqref{zeropvanish} has been proved for the CDvG-Grushevsky
(CDvG-G) ansatz at genus $3$ \cite{Cacciatori:2008ay} and $4$
\cite{Grushevsky:2008zm}. Remarkably, for genus $4$, $A_0$
corresponds to a non-zero Siegel modular form of weight $8$ (the
Schottky form), which vanishes only on the locus of Jacobians of
Riemann surfaces. A strong argument for the identities
\eqref{twopvanish} and \eqref{threepvanish} to hold on the
hyperelliptic locus for any genus has been given by Morozov in
\cite{Morozov:2008xd}, whereas in \cite{Grushevsky:2008qp}
Grushevsky and Salvati Manni proved \eqref{twopvanish} for genus
$3$.

\medskip

 In this paper, we will prove that
\eqref{threepvanish} does not hold for any non-hyperelliptic Riemann
surface of genus $3$. More precisely, one of the main results of the
paper is the following theorem
\medskip
\begin{theorem}\label{th:main} Let $C$ be a Riemann surface of genus
three. Then, $A_3(a,b,c)=0$ for all $a,b,c\in C$ if and only if $C$
is hyperelliptic.\end{theorem}
\medskip
In particular, we will prove that the remarkable identity
\begin{equation}\label{A3formula}A_3(p_1,p_2,p_3)d\mu_{Bos}=\frac{c}{c_3}\det\omega_i(p_j)\prod_{i\le
j}d\tau_{ij}\ ,
\end{equation}
where $c_3:=2^6\pi^{18}$ and $c\in\CC$ is a non-vanishing constant,
holds for genus $3$, so giving a close expression for $A_3$.
Moreover, we discuss the possibility of a further non-vanishing
contribution to the three-point function coming from the connected
part of the chiral amplitude and give some arguments suggesting that
the natural candidate for such a term should exhibit the same
structure of the right-hand side of \eqref{A3formula}. In this
respect, \eqref{A3formula} can be interpreted as the statement of
the non-renormalization of the three-point function at three-loop.

The paper is organized as follows. In section \ref{s:mathback} we
introduce the mathematical background on Riemann surfaces and theta
functions needed for the later construction. In section
\ref{s:vanish} we give a necessary and sufficient condition for
$A_3(a,b,c)$ to vanish identically and we find that such a condition
is trivially satisfied for $g=2$, whereas, for genus $3$, it is
fulfilled only for hyperelliptic surfaces. In section
\ref{s:goodform}, a strikingly simple formula for $A(a,b,c)$ is
provided for genus $3$, which does not include any summation over
spin characteristics; equation \eqref{A3formula} is an immediate
consequence of such a formula. In section \ref{s:discussion}, we
discuss how the non-renormalization theorems could be implemented in
the framework of the CDvG-G ans\"atze for the chiral superstring
measures in view of our results.

\section{Theta functions and Riemann surfaces}\label{s:mathback}

In this section, we provide the basic background on theta functions
and Riemann surfaces necessary for the subsequent derivations. We
refer to \cite{Mumford:1983,Fay:1973,Farkas:1992} for proofs and
further details.

\medskip

Let $\Sieg_g$ denote the Siegel upper half-space, i.e. the space of
$g\times g$ complex symmetric matrices with positive definite
imaginary part
$$\Sieg_g:=\{\tau\in M_{g\times g}(\CC)\mid \tp\tau=\tau\,
,\im\tau>0 \}\ .
$$
Let $\Sp(2g,\ZZ)$ be the symplectic modular group, i.e. the group of
$2g\times 2g$ complex matrices $M:=\bigl(\begin{smallmatrix}A &B\\
C & D\end{smallmatrix}\bigr)$, where $A,B,C,D$ are $g\times g$
blocks satisfying
$$\tp A C=\tp C A\ ,\quad \tp BD=\tp D B\ ,\quad \tp
D A-\tp B C=\II_g\ .
$$
Let us define the action of $\Sp(2g,\ZZ)$ on $\CC^g\times\Sieg_g$ by
\begin{equation}\label{modull}
(M\cdot z,M\cdot\tau):=\bigl(\tp(C\tau+D)^{-1}z,(A\tau+B)(C\tau+D)^{-1}\bigl)\ ,\end{equation} where $M\equiv\bigl(\begin{smallmatrix}A &B\\
C & D\end{smallmatrix}\bigr)\in\Sp(2g,\ZZ)$ and
$(z,\tau)\in\CC^g\times\Sieg_g$.

\medskip

Let $\ZZ_2:=\ZZ/(2\ZZ)$ be the additive group with elements
$\{0,1\}$. For each $\delta',\delta'' \in\ZZ_2^{g}$, the theta
function $\theta[\delta]\equiv\theta\spin{\delta'}{\delta''}\colon
\CC^g\times\Sieg_g\to \CC$ with characteristics $[\delta]
\equiv\spin{\delta'}{\delta''}$
 is defined by
$$\theta[\delta](z,\tau):=\sum_{k\in\ZZ^g}
\exp{\pi
i\Bigl[\tp\Bigl(k+\frac{\delta'}{2}\Bigr)\tau\Bigl(k+\frac{\delta'}{2}\Bigr)+
2\tp\Bigl(k+\frac{\delta'}{2}\Bigr)\Bigl(z+\frac{\delta''}{2}
\Bigr)\Bigr]\ ,}
$$
where $(z,\tau)\in\CC^g\times\Sieg_g$. For each fixed $\tau$,
$\theta[\delta](z,\tau)$ is an even or odd function on $\CC^g$
depending whether $(-1)^{\delta'\cdot\delta''}$ is $+1$ or $-1$,
respectively. Correspondingly, there are $2^{g-1}(2^g+1)$ even and
$2^{g-1}(2^g-1)$ odd theta characteristics. Under translations
$z\mapsto z+\lambda$, $z\in\CC^g$, $\lambda
\in\ZZ^g+\tau\ZZ^g\subset\CC^g$, theta functions get multiplied by a
nowhere vanishing factor
$$\theta \spin{\delta'}{\delta''}\left(z+n+\tau m,\tau\right)=
e^{-\pi i \tp{m}\tau m-2\pi i\tp{m}z+\pi
i(\tp{\delta'}n-\tp{\delta''}m)}\theta\spin{\delta'}{\delta''}\left(z,\tau\right)
\ ,
$$
$m,n\in\ZZ^g$. It follows that, for any fixed $\tau$, the theta
functions can be seen as sections of line bundles on the complex
torus $A_\tau:=\CC^g/(\ZZ^g+\tau\ZZ^g)$, with a well defined divisor
on $A_\tau$. We denote by $\Theta$ the divisor of
$\theta(z)\equiv\theta[0](z,\tau)\equiv\theta\spin{0}{0}(z,\tau)$
and by $\Sing\Theta$ its singular locus, i.e. the locus of points at
which $\theta(z)$ and all its first partial derivatives vanish.

The second order theta functions are defined by
$$\Theta[\epsilon](z,\tau):=\theta\spin{\epsilon }{0}(2z,2\tau)\ ,
$$
for all $\epsilon\in\ZZ^g$. They are a basis for
$H^0(A_\tau,\OO(2\Theta))$ and are related to the first order theta
functions by the Riemann bilinear identities
\begin{equation}\label{RiemBil}\theta\spin{\epsilon}{\delta
}(z_1+z_2,\tau)\,\theta\spin{\epsilon}{\delta
}(z_1-z_2,\tau)=\sum_{\sigma\in\ZZ^{g}}(-1)^{\delta\cdot\sigma}\Theta[\sigma](z_1,\tau)\,\Theta[\sigma+\epsilon
](z_2,\tau)\ ,
\end{equation}
for all $z_1,z_2\in\CC^g$, $\epsilon ,\delta \in\ZZ_2^g$.

\medskip

Let us define the action of $\Sp(2g,\ZZ)$ on $\ZZ_2^{2g}$ by
 \begin{equation}
\label{modulcharac}M\cdot\delta\equiv M\cdot
\Biggl(\begin{matrix}\delta' \\
\delta''\end{matrix}\Biggr):=\Biggl(\begin{matrix}D& -C\\ -B
&A\end{matrix}\Biggr)\Biggl(\begin{matrix}\delta' \\
\delta''\end{matrix}\Biggr)+\Biggl(\begin{matrix}\diag (C\tp
D)\\\diag (A\tp B)\end{matrix}\Biggr)\mod 2\ .
\end{equation}
Theta characteristics are invariant under the action of the subgroup
$\Gamma(2)\subset\Sp(2g,\ZZ)$, where
$$\Gamma(n):=\{M\in\Sp(2g,\ZZ)\mid M=\II_{2g}\mod n\}\ ,
$$ is the subgroup of elements of $\Sp(2g,\ZZ)$ congruent to the $2g\times 2g$
identity matrix mod $n$. The action of $\Sp(2g,\ZZ)$ on $\ZZ_2^{2g}$
factorizes through the action of
$\Sp(2g,\ZZ_2)\equiv\Sp(2g,\ZZ)/\Gamma(2)$. Symplectic
transformations preserve the parity of the characteristics and, for
any two $\delta,\epsilon\in\ZZ_2^{2g}$ of the same parity, there
exists $M\in\Sp(2g,\ZZ_2)$ such that $\epsilon=M\cdot\delta$.

A (Siegel) modular form $f$ of weight $k\in\ZZ$ for a subgroup
$\Gamma\in\Sp(2g,\ZZ)$ is a holomorphic function on $\Sieg_g$ such
that
$$f(M\cdot \tau)=\det(C\tau+D)^kf(\tau)\ ,
$$
for all $M\in\Gamma$. A condition of regularity is also required for
$g=1$, but it is automatically satisfied for $g>1$.

\medskip
 The modular transformation of the theta function is given
by
\begin{equation}\label{modultheta}
\theta[M\cdot\delta](M\cdot z,M\cdot \tau)=\kappa(M)\det
(C\tau+D)^\frac{1}{2} e^{\pi i[\phi[\delta](M)+\tp
z(C\tau+D)^{-1}Cz]}\theta[\delta](z,\tau)\ ,\end{equation} where
$\kappa(M)$ is an eighth root of $1$ depending on $M$ and
$$4\phi\spin{\delta'}{\delta''}(M):=(\tp \delta' \ \tp \delta'')\Biggl(\begin{matrix} -\tp BD & \tp BC\\ \tp BC & -\tp
AC\end{matrix}\Biggr)\Biggl(\begin{matrix}\delta'\\ \delta''
\end{matrix}\Biggr)+2\diag(A\tp B)\cdot (D\delta'-C\delta'')\ .$$
Powers of theta constants
$\theta[\delta](\tau)\equiv\theta[\delta](0,\tau)$ are the basic
building blocks for modular forms, at least for low genera.

\medskip

Let $C$ be a Riemann surface of genus $g>1$. The choice of a marking
for $C$ provides a set of generators
$\{\alpha_1,\ldots,\alpha_g,\beta_1,\ldots,\beta_g\}$ for the first
homology group $H_1(C,\ZZ)$ on $C$, with symplectic intersection
matrix, that is
\begin{equation}\label{sympl}\alpha_i\cdot\alpha_j=0=\beta_i\cdot\beta_j\ ,\qquad
\alpha_i\cdot\beta_j=\delta_{ij}\ ,
\end{equation}
for all $i,j=1,\ldots,g$. The choice of such generators canonically
determines a basis $\{\omega_1,\ldots,\omega_g\}$ for the space
$H^0(K_C)$ of holomorphic $1$-differentials on $C$, with normalized
$\alpha$-periods
$$\oint_{\alpha_i}\omega_j=\delta_{ij}\ ,$$
for all $i,j=1,\ldots,g$. The $\beta$-periods define the Riemann
period matrix of the Riemann surface $C$
$$\tau_{ij}\equiv\oint_{\beta_i}\omega_j\ ,
$$
which is symmetric and with positive-definite imaginary part, so
that $\tau\in\Sieg_g$. By Torelli's theorem, the complex structure
of $C$ is completely determined by giving its Riemann period matrix.

By the conditions \eqref{sympl}, a general change of marking of $C$
corresponds to a symplectic transformation on the set of generators
of $H_1(C,\ZZ)$
\begin{equation}\label{modulhomol}
\Biggl(\begin{matrix}\alpha\\
\beta\end{matrix}\Biggr)\mapsto 
\Biggl(\begin{matrix}\tilde\alpha\\
\tilde\beta\end{matrix}\Biggr):=\Biggl(\begin{matrix}D & C \\ B &
A\end{matrix}\Biggr) \Biggl(\begin{matrix}\alpha\\
\beta\end{matrix}\Biggr)\ ,\qquad\qquad
M\equiv\Biggl(\begin{matrix}A & B
\\ C & D\end{matrix}\Biggr)\in \Sp(2g,\ZZ) \ , \end{equation} under which
\begin{equation}\label{modulbasis}
(\omega_1,\ldots,\omega_g)\mapsto(\tilde\omega_1,\ldots,\tilde\omega_g):=(\omega_1,\ldots,\omega_g)(C\tau+D)^{-1}\
,
\end{equation}
whereas $\tau\mapsto\tilde\tau:=M\cdot\tau$ transforms as in
\eqref{modull}.

\medskip

 The
complex torus $J_C:=\CC^g/(\ZZ^g+\tau\ZZ^g)$ associated to the
Riemann period matrix of $C$ is called the Jacobian torus of $C$.
For a fixed base-point $p_0\in C$, let $I\colon C\to J_C$ denote the
Abel-Jacobi map, defined by
$$p\mapsto
I(p):=\tp\Bigl(\int_{p_0}^p\omega_1,\ldots,\int_{p_0}^p\omega_g\Bigr)\in
J_C\ .
$$
Note that different choices of the path of integration from $p_0$ to
$p$ correspond, by the formula above, to points in $\CC^g$ differing
by elements in the lattice $\ZZ^g+\tau\ZZ^g$, so that $I$ is
well-defined only on $\CC^g/(\ZZ^g+\tau\ZZ^g)$. The Abel-Jacobi map
extends to a map from the Abelian group of divisors on $C$ to $J_C$
by
$$I\bigl(\smsum\nolimits_i p_i-\smsum\nolimits_i q_i\bigl):=\smsum\nolimits_i I(p_i)-\smsum\nolimits_i I(q_i)\ .$$
Such a map is independent of the base point $p_0$ when restricted to
zero degree divisors. In the following, when no confusion is
possible, we will identify such zero degree divisors with their
image in $J_C$ through $I$. In particular, we will omit $I$ when
considering the theta functions on the Jacobian evaluated at (the
image of) some zero degree divisor on $C$. Furthermore, the argument
$\tau$ for theta functions associated to a marked Riemann surface
will be understood.

\medskip

Given a Riemann surface $C$ with marking, one can canonically
associate to each theta characteristic $\delta\in\ZZ_2^{2g}$ a spin
structure, correspondent to a line bundle $L_\delta$ on $C$ such
that $L_\delta^2\cong K_C$, with $K_C$ the canonical line bundle on
$C$. Such a correspondence can be defined as follows. Let $\delta$
be a non-singular theta characteristic (that is, such that at least
one among $\theta[\delta](z)$ and its first partial derivatives does
not vanish at $z=0$) and, for an arbitrary $y\in C$, set
$f_{\delta,y}(x):=\theta[\delta](x-y)$. By the Riemann vanishing
theorem \cite{Farkas:1992}, the divisor $2(f_{\delta,y})-2y$ is
linearly equivalent to the canonical divisor, so that
$(f_{\delta,y})-y$ defines the divisor class of a spin bundle, that
we denote by $L_\delta$. It can be proved that such a divisor class
$[(f_{\delta,y})-y]$, and thus also $L_\delta$, is independent of
$y\in C$, so that for each marked Riemann surface we have a
correspondence $\delta\mapsto L_\delta$. By \eqref{modultheta},
under a change of marking \eqref{modulhomol}, such a correspondence
transforms into the map $\delta\mapsto \tilde L_\delta$, where
\begin{equation}\label{modulspin}\tilde L_{M\cdot\,\delta}=L_{\delta}\
.\end{equation}

Fix a non-singular odd spin structure $\nu\in\ZZ_2^{2g}$ and
consider the holomorphic $1$-differential
$$\sum_{i=1}^g\left.\frac{\partial\theta[\nu](z)}{\partial
z_i}\right\rvert_{z=0}\omega_i\ .
$$
It can be proved that such a $1$-differential has $g-1$ double
zeroes and corresponds to the square $h_\nu^2$ of a holomorphic
section of the line bundle $L_\nu$.

Let us define the prime form by
$$E(a,b):=\frac{\theta[\nu](b-a)}{h_\nu(a)h_\nu(b)}\ ,
$$
$a,b\in C$, for an arbitrary non-singular odd spin-structure $\nu$.
The prime form is a section of a line bundle on $C\times C$, it is
antisymmetric in its arguments and vanishes only on the diagonal
$a=b$. Furthermore, it does not depend on the choice of $\nu$.

For each non-singular even characteristic $\delta\in\ZZ^{2g}$, the
Szeg\"o kernel is defined by
$$S_\delta(a,b)\equiv S(a,b;L_\delta):=\frac{\theta[\delta](a-b)}{\theta[\delta](0)\,E(a,b)}\
.
$$ For each fixed $b\in C$, $S_\delta(a,b)$ is the unique meromorphic
section of $L_\delta$ with a single pole of residue $-1$ at $b$ and
holomorphic elsewhere. Such a characterization implies that, for a
fixed spin bundle $L$, $S(a,b;L)$ is independent of the marking. It
follows that, under a change of marking corresponding to
\eqref{modulhomol}, by \eqref{modulspin} we have
\begin{equation}\label{modulSzego}
\tilde S_{M\cdot\,\delta}(a,b)\equiv S(a,b;\tilde
L_{M\cdot\,\delta})= S(a,b;L_{\delta}) =S_{\delta}(a,b)\
,\end{equation} with $M\cdot\delta$ given by \eqref{modulcharac}.

Finally, we denote by $$\omega_{a-b}(x):=\frac{\partial}{\partial
x}\log\frac{E(x,a)}{E(x,b)}\ ,$$ $a,b,x\in C$, the Abelian
$1$-differential of the second kind with single poles on $a$ and $b$
with residue $+1$ and $-1$, respectively, holomorphic on
$C\setminus\{a,b\}$, and with vanishing $\alpha$-periods.


\section{Proof of theorem \ref{th:main}}\label{s:vanish}

Let $\Xi[\delta](\tau)$ be the modular form of weight $8$ for
$\Gamma(2)\subset\Sp(2g,\ZZ)$ defined in \cite{Cacciatori:2008ay}
for genus $3$ and in \cite{Grushevsky:2008zm} for arbitrary genus.
It satisfies the property
\begin{equation}\label{modulXi}\Xi[M\cdot\delta](M\cdot\tau)=\det
(C\tau+D)^8\,\Xi[\delta](\tau)\ ,
\end{equation}
for an arbitrary $M\in\Sp(2g,\ZZ)$.

In terms of $\Xi[\delta]$ and the Szeg\"o kernel, one can define the
sections $A_2(a,b)$ on $C\times C$ and $A_3(a,b,c)$ on $C\times
C\times C$ by \eqref{twopvanish} and \eqref{threepvanish},
respectively. In \cite{Grushevsky:2008qp}, it has been proved that
$A_2(a,b)=0$ for all $a,b\in C$, where $C$ is an arbitrary Riemann
surface of genus $2$ or $3$. It is useful to recall the main points
of such a derivation.

Let $C$ be a marked Riemann surface of genus $g>1$ and let $\tau$ be
its Riemann period matrix. Define the function $X\colon\CC^g\to\CC$
by
\begin{equation}\label{Xsect}X(z):=\sum_{\delta\text{ even}}\Xi[\delta]\frac{\theta[\delta
](z,\tau)^2}{\theta[\delta ](0,\tau)^2}\ ,
\end{equation}
$z\in \CC^g$, corresponding to a section of $|2\Theta|$ on the
Jacobian $J_C$. The restriction of such a section to
$$C-C:=\{(a-b)\in J_C, a,b\in C\}\subset J_C\ ,
$$
is related to $A_2(a,b)$ by
$$A_2(a,b)=\frac{X(a-b)}{E(a,b)^2}\ .
$$
Thus, $A_2(a,b)=0$ for all $a,b\in C$ if and only if the restriction
of $X$ to $C-C$ vanishes identically. On the other hand, as
conjectured in \cite{VanGeemen:1986} and proved in
\cite{Welters:1986}, the space of sections of $|2\Theta|$ vanishing
on $C-C$ is
$$\Gamma_{00}:=\{f\in H^0(J_C,\OO(2\Theta))\mid
f(0)=0=\partial_i\partial_j f(0),\ i,j=1,\ldots,g\}\ ,
$$
so that the following theorem follows.
\medskip
\begin{theorem}[Grushevsky, Salvati Manni \cite{Grushevsky:2008qp}]\label{th:GrSalMan} The function $A_2(a,b)=0$ for all $a,b\in C$
if and only if $X\in\Gamma_{00}$.
\end{theorem}
\medskip
For genus $g>1$, the dimension of $\Gamma_{00}$ is $2^g-1-g(g+1)/2$;
in particular, $\dim\Gamma_{00}=0$ for $g=2$ and $\dim\Gamma_{00}=1$
for $g=3$. By applying the Riemann bilinear relations, $X(z)$ can be
expressed in terms of second order theta functions. In
\cite{Grushevsky:2008qp}, such an expression is used to show that
the function $X(z)$ vanishes identically on $\CC^g$ for $g=2$,
 whereas for $g=3$ $X(z)$ is a generator of
$\Gamma_{00}$. By theorem \ref{th:GrSalMan}, this proves that
$A_2(a,b)\equiv 0$ for $g=2,3$.

\medskip

Let us show that, for an arbitrary marked Riemann surface $C$ of
genus $g>1$, also $A_3(a,b,c)$ can be expressed in terms of $X(z)$.
The starting point is the following identity
\begin{equation}\label{FaySzego}\frac{S_\delta(c,a)S_\delta(b,c)}{S_\delta(a,b)}=\omega_{a-b}(c)+\sum_{i=1}^g
\left.\frac{\partial\log\theta[\delta](z)}{\partial z_i}\right\vert
_{z=a-b}\,\omega_i(c)\ , \end{equation} $a,b,c\in C$, which is an
immediate consequence of \cite{Fay:1973}, formula 38 page 25. It is
instructive to derive such an identity from the famous Fay's
trisecant identity \cite{Fay:1973}, written in the form
\begin{equation}\label{Fayid}\frac{\theta[\delta](a+c-b-d)E(a,c)E(b,d)}
{\theta[\delta](0)E(a,b)E(a,d)E(c,b)E(c,d)}=S_\delta(a,d)S_\delta(c,b)-
S_\delta(a,b)S_\delta(c,d) \ ,
\end{equation}
which holds for arbitrary $a,b,c,d\in C$ and for each non-singular
even spin structure $\delta$. By comparing the Laurent expansion of
both sides of \eqref{Fayid} in the limit $d\to c$, with respect to
some local coordinate centered in $c$, we obtain an infinite tower
of (possibly trivial) identities, one for each order in $(d-c)$.
Using $E(c,d)^{-1}=(d-c)^{-1}(1+O(d-c)^2)$, it is easy to check that
the first non-trivial identity is obtained at $O(1)$
$$-\frac{\theta[\delta](a-b)}{\theta[\delta](0)E(a,b)}\left.\frac{d}{d
x}
\Bigl(\log\theta[\delta](a+c-b-x)+\log\frac{E(b,x)}{E(a,x)}\Bigr)\right|_{x=c}\!\!\!=S_\delta(a,c)S_\delta(c,b)\
,
$$
and \eqref{FaySzego} follows immediately. (Note that second term on
the RHS of \eqref{Fayid} is odd in
$(d-c)$, so that it does not contribute to $O(1)$.) 

By multiplying both sides of \eqref{FaySzego} by
$\Xi[\delta]S_{\delta}(a,b)^2$ and summing over all the even spin
structures, we obtain
\begin{multline*}\sum_{\delta\text{ even}}\Xi[\delta]S_\delta(c,a)S_\delta(b,c)S_\delta(a,b)=
\omega_{a-b}(c)\sum_{\delta\text{ even}}\Xi[\delta]S_\delta(a,b)^2\\
+ \frac{1}{E(a,b)^2}\sum_{\delta\text{ even}}\frac{\Xi[\delta]}
{\theta[\delta](0)^2}\theta[\delta](a-b)
\sum_{i=1}^g\left.\frac{\partial\theta[\delta](z)}{\partial
z_i}\right\vert_{z=a-b}\omega_i(c)\ .
\end{multline*}
By \eqref{Xsect}, such an identity can be written as
\begin{equation}\label{threepoint}A_3(a,b,c)=
\frac{1}{E(a,b)^2}\Bigl[\omega_{a-b}(c)X(a-b)+
\frac{1}{2}\sum_{i=1}^g\left.\frac{\partial X(z)}{\partial
z_i}\right\vert_{z=a-b}\omega_i(c)\Bigr]\ .\end{equation} Since
$\omega_{a-b},\omega_1,\ldots,\omega_g$ are linearly independent,
the condition that $A_3(a,b,c)=0$ for all $a,b,c\in C$, is
equivalent to the $g+1$ conditions
\begin{align*}&X(a-b)=0\ ,\\ &\left.\frac{\partial
X(z)}{\partial z_i}\right\vert_{z=a-b}=0\ ,\qquad\qquad
i=1,\ldots,g\ ,
\end{align*} to hold for all $a,b\in C$. Note that the first
condition is equivalent to $A_2(a,b)=0$. It is natural to define the
following subspace of $\Gamma_{00}$
$$\Gamma^{(2)}_{00}:=\{f\in \Gamma_{00}\mid \mult_{a-b}(f)\ge 2,\ \forall a,b\in C\}\
.
$$
We have proved the following theorem.
\medskip
\begin{theorem}\label{th:threevanish} For an arbitrary Riemann surface $C$ of genus $g>1$,
$A_3(a,b,c)=0$ for all $a,b,c\in C$  if and only if
$X\in\Gamma^{(2)}_{00}$.
\end{theorem}
\medskip
The space $\Gamma_{00}^{(2)}$ and other remarkable subspaces of
$\Gamma_{00}$ have been extensively studied in the last few years,
in particular because of their relationship with the geometry of the
moduli space $\mathcal{S}U_C(2,K)$ of semi-stable bundles of rank 2
with fixed canonical determinant on a smooth projective curve $C$
\cite{Oxbury:1998,Pauly:2001,Izadi:2001}. There is a simple
procedure to construct elements of $\Gamma_{00}^{(2)}$. Let
$e\in\Sing\Theta$ be a point of the singular locus of the theta
function, i.e. such that $\theta(z)$ and all its first derivatives
vanish at $z=e$; note that, by the parity of the theta function,
also $-e\in\Sing\Theta$. By the Riemann singularity theorem
\cite{Farkas:1992}, $\theta(a-b+e)=0$ for all $a,b\in C$. It follows
immediately that
$$F_e(z):=\theta(z+e)\theta(z-e)=\sum_{\sigma\in\ZZ_2^g}\Theta[\sigma](e)\,\Theta[\sigma](z)\ ,
$$
is an element of $\Gamma^{(2)}_{00}$. In \cite{Pauly:2001} (Theorem
1.1), it has been proved that, for an arbitrary non-hyperelliptic
Riemann surface $C$, $\Gamma^{(2)}_{00}$ is generated by the
sections $F_e(z)$ as $e$ varies in $\Sing\Theta$, that is
$\Gamma^{(2)}_{00}=\langle F_e\rangle_{e\in \Sing\Theta}$.

\medskip

Let us consider the consequences of such results for low genera. For
genus $2$, the identity $X\equiv 0$ proved in
\cite{Grushevsky:2008qp}, together with theorem
\ref{th:threevanish}, implies that $A_3(a,b,c)\equiv0$, thus
reobtaining the result of \cite{D'Hoker:2005jc}.

\medskip
 For genus $3$, in the
non-hyperelliptic case, $\Sing\Theta$ is empty, so that
$\dim\Gamma_{00}^{(2)}=0$. On the other hand, in
\cite{Grushevsky:2008qp} it has been proved that, in this case,
$X\neq 0$, so that we conclude that $A_3(a,b,c)$ does not vanish
identically on $C\times C\times C$.

The hyperelliptic curves are not considered in \cite{Pauly:2001}. In
this case, one has just the weaker inclusion $\langle
F_e\rangle_{e\in \Sing\Theta}\subseteq \Gamma^{(2)}_{00}$. On the
other hand, if $C$ is hyperelliptic of genus $3$,
$\Sing\Theta\subset J_C$ consists of a unique point of order $2$,
corresponding to a singular even spin-structure $\delta_{sing}$. It
follows that, in this case, $\Gamma_{00}^{(2)}$ has at least one
non-trivial section, namely
$$F_{\delta_{sing}}(z):=\theta[\delta_{sing}](z)^2\ .
$$
Since $\Gamma_{00}$ is $1$-dimensional and contains
$\Gamma_{00}^{(2)}$ as a subspace, we get
$\Gamma_{00}=\Gamma_{00}^{(2)}$ so that $X\in\Gamma_{00}^{(2)}$ and
$A_3(a,b,c)=0$ for all $a,b,c\in C$, as suggested by the arguments
in \cite{Morozov:2008xd}. This concludes the proof of theorem
\ref{th:main}.

\medskip
More generally, for genus $g\ge 3$, the space $\langle
F_e\rangle_{e\in \Sing\Theta}$ has dimension
$2^g-\sum_{i=0}^3\binom{g}{i}$, which, for non-hyperelliptic
surfaces, corresponds to the dimension of $\Gamma_{00}^{(2)}$ . In
particular, for a non-hyperelliptic surface of genus $4$,
$\dim\Gamma_{00}^{(2)}=1$. In this case, $\Sing\Theta$ has only two
(possibly coincident) points $\pm e$ and the generator of
$\Gamma_{00}^{(2)}$ is $\theta(z+e)\theta(z-e)$. It would be
interesting (but probably highly non-trivial) to check whether $X$
is proportional to such a section in this case.

\section{A simple expression for $A_3(a,b,c)$}\label{s:goodform}

Let us show that $A_3(a,b,c)$ admits the  alternative expression
\eqref{A3formula} that will be useful in the following.

First of all, note that, if $A_2(a,b)$ vanishes identically on a
Riemann surface $C$ of genus $g>1$, then $A_3(a,b,c)$ is a
holomorphic $1$-differential in each variable. Furthermore, it is
anti-symmetric under permutation of such variables and, in
particular, it must vanish on the diagonals of $C\times C\times C$.
For $g=3$, this is enough to conclude that
$$A_3(p_1,p_2,p_3)\equiv\sum_{\delta\text{ even}}\Xi[\delta]S_\delta(p_1,p_2)S_\delta(p_2,p_3)S_\delta(p_3,p_1)
=f\det\omega_i(p_j)\ ,
$$
$p_1,p_2,p_3\in C$, for some holomorphic function $f$, independent
of $p_1,p_2,p_3$, on the Teichm\"uller space of genus $3$ Riemann
surfaces with marking. Under a change of marking, corresponding to a
transformation \eqref{modulhomol} for some $M\equiv\bigl(\begin{smallmatrix}A & B\\
C & D\end{smallmatrix}\bigr)\in\Sp(2g,\ZZ)$, by
\eqref{modulbasis}\eqref{modulSzego}\eqref{modulXi} we get
\begin{align*}
A_3(p_1,p_2,p_3)&\mapsto\det(C\tau+D)^8A_3(p_1,p_2,p_3)\ ,\\
\det \omega_i(p_j)&\mapsto\det(C\tau+D)^{-1}\det\omega_i(p_j)\
.\end{align*} It follows that $f$ transforms as
$$f\mapsto\det(C\tau+D)^9f\ ,
$$
and thus it corresponds to a Teichm\"uller modular form of weight
$9$ and degree $3$ \cite{Ichikawa:1995}. In general, a Teichm\"uller
modular form of weight $d\ge 0$ and degree $g$ is defined as a
holomorphic section of $\lambda_1^{\otimes d}$ on the moduli space
$\M_g$ of genus $g$ Riemann surfaces; here, $\lambda_1$ is the line
bundle whose fiber at the point corresponding to the surface $C$ is
$\wedge^g H^0(K_C)$. By \cite{Ichikawa:1995}, there is a unique (up
to a constant), holomorphic section $\mu_{3,9}$ of
$\lambda_1^{\otimes 9}$ on $\M_3$, so that
$$f=c\,\mu_{3,9}\ ,
$$
for some $c\in\CC$; theorem \ref{th:main} implies that $c\neq 0$.
The section $\mu_{3,9}$ vanishes only on the hyperelliptic locus,
consistently with theorem \ref{th:main}, and its square
$(\mu_{3,9})^2$ corresponds to the (Siegel) modular form
$$\Psi_{18}(\tau):=\prod_{\delta\text{ even}}\theta[\delta](0,\tau)\
,
$$
of weight $18$. (More precisely, $(\mu_{3,9})^2$ is the image of
$\Psi_{18}$ under the homomorphism, induced by the Torelli map,
mapping Siegel modular forms to Teichm\"uller modular forms
\cite{Ichikawa:1995}; for such a reason, $\mu_{3,9}$ itself is often
identified with the parabolic form $\Psi_9$, that is a holomorphic
square root of $\Psi_{18}$).

Furthermore, $\mu_{3,9}$ also appears in the explicit formula for
the chiral bosonic string measure for genus $3$
\cite{Belavin:1986tv}
$$d\mu_{Bos}=\frac{1}{c_3}\frac{\prod_{i\le j}d\tau_{ij}}{\mu_{3,9}}\ ,
$$
with $c_3=2^6\pi^{18}$ \cite{D'Hoker:2004ce}, and the identity
\eqref{A3formula} follows.

It would be interesting to compute exactly the constant $c$. This
may be done, for example, by using the factorization properties of
string amplitudes under degeneration limits.

\section{Non-renormalization theorems and chiral measure ans\"atze}\label{s:discussion}

Theorem \ref{th:main} shows that the CDvG-G ansatz and the
assumption \eqref{connvanish} are not compatible with the
non-renormalization theorems at $3$-loops. It is necessary,
therefore, to discuss the validity of the assumptions leading to
such a result.

For genus $3$, upon assuming the form \eqref{threeloopmeas}, the
chiral superstring measure is completely determined by the
constraints related to holomorphicity, modular invariance and
factorization. Even though there are no strong first-principle
arguments suggesting that \eqref{threeloopmeas} holds for genus
higher than $2$, a posteriori there are several hints that the
ansatz should be correct, at least for low genera: the uniqueness of
the genus $3$ solution for the constraints, the existence of general
well-defined solutions at least up to genus $5$, the
non-renormalization of the cosmological constant at $3$ and $4$
loops (a result that, as noted in the introduction, is independent
of the assumption \eqref{connvanish}).

The evidence for the assumption \eqref{connvanish} is much weaker.
In fact, it does not hold, for example, for the $2$-loop
contribution to the $4$-point function \cite{D'Hoker:2005jc}.
Therefore, it is reasonable to consider the possibility of a
non-vanishing contribution from the connected part of the chiral
amplitude. Some hints on the possible structure of such a
contribution can be obtained analyzing the $4$-point function at $2$
loops. In this case, the connected part of the chiral amplitude
gives two non-vanishing terms, both corresponding to Wick
contractions between the vertex operators and the stress-energy
tensor. One of such contributions exactly cancels the disconnected
part of the amplitude. The other one has, very schematically, the
following structure
\begin{equation}\label{A4formula}A_4(p_1,p_2,p_3,p_4)d\mu_{Bos}=
\sum_{I_1\sqcup I_2=\{1,2,3,4\}}\!
K_{I_1}\,\det_{\substack{i=1,2\\j\in I_1}}\omega_i(p_j)
\det_{\substack{i=1,2\\j\in I_2}}\omega_i(p_j)\prod_{i\le
j}d\tau_{ij}\ ,
\end{equation}
where $K_{I_1}$ is a kinematical factor and the sum is over all the
possible ways of splitting the set $\{1,2,3,4\}$ into the disjoint
union of sets $I_1$ and $I_2$ of two elements. Hence, it seems
reasonable for a non-vanishing $3$-loop contribution corresponding
to Wick contractions between vertex operators and stress tensor to
exhibit a structure analogous to the right-hand side of
\eqref{A4formula}.

On the other hand, the most natural generalization of
\eqref{A4formula} to the case of a $3$-loop contribution to the
$3$-point function, satisfying the fundamental consistency
constraints (modular weight $5$ and conformal weight $1$ in each
variable), is precisely the right-hand side of \eqref{A3formula}
(times a kinematical factor). In this respect, note that the
requirements we are imposing on the possible structure of such a
contribution are very restrictive. In fact, no consistent
generalization of \eqref{A4formula} can be defined for the $2$-loop
contribution to the $3$-point function or for the $2$- and $3$-loop
contributions to the $2$-point function. The fact that, for such
amplitudes, the disconnected part vanishes separately enforces the
validity of our analysis. Remarkably, a structure similar to
\eqref{A3formula} has been proposed in \cite{Matone:2005vm} for the
higher loop contributions to the $4$-point function.

Thus, by the identity \eqref{A3formula} and by the arguments above,
it is reasonable to conjecture that the contributions from the
connected chiral amplitude could exactly cancel the disconnected
part, thus giving the expected non-renormalization theorems. It
would be very interesting to check such a conjecture by explicitly
computing some of the terms coming from the relevant Wick
contractions. Unfortunately, even though such a computation should
be considerably simpler than a complete first-principles derivation
of the amplitudes (for example, we are neglecting the Wick
contractions between supercurrents and vertex operators, together
with a huge number of subtleties already emerging at genus $2$), it
is not clear, at the moment, how it should be performed.

\bigskip

\section*{Acknowledgements}
We are grateful to Riccardo Salvati Manni for several interesting
discussions, to Emma Previato for clarifying comments about the
space $\Gamma_{00}^{(2)}$ and to Duong Phong and Samuel Grushevsky
for helpful discussions on superstring amplitudes.

\end{document}